\documentclass[aps,prl,twocolumn,showpacs]{revtex4}
\usepackage{graphicx}
\usepackage{amssymb}
\usepackage{color}
\begin{document}


\title{Plaquette Order in the $J_1$-$J_2$-$J_3$ model: a series expansion
analysis}



\author{Marcelo Arlego}

\affiliation{Departamento de F\'isica, Universidad Nacional de La Plata, C.C. 67,
1900 La Plata, Argentina.}

\email[E-mail: ]{m.arlego@fisica.unlp.edu.ar}

\author{Wolfram Brenig}

\affiliation{Institut f\"ur Theoretische Physik, Technische
Universit\"at Braunschweig, 38106 Braunschweig, Germany}


\date{\today}

\begin{abstract}
Series expansion based on the flow equation method is employed to study the
zero temperature properties of the spin$-1/2$ $J_1$-$J_2$-$J_3$
antiferromagnet in two dimensions. Starting from the exact limit of
decoupled plaquettes in a particular \emph{generalized} $J_1$-$J_2$-$J_3$
model we analyze the evolution of the ground state energy and the
elementary triplet excitations in powers of all three inter-plaquette
couplings up to fifth order. We find the plaquette phase to remain stable
over a wide range of exchange couplings and to connect adiabatically up
to the case of the plain $J_1$-$J_2$-$J_3$ model, however not to
the $J_1$-$J_2$ model at $J_3=0$. Besides confirming the existence of such
a phase, recently predicted by Mambrini, {\em et al.} [Phys. Rev. B
\textbf{74}, 144422 (2006)], we estimate its extent by Dlog-Pad\'e analysis
of the critical lines that result from closure of the triplet gap.
\end{abstract}



\maketitle


\section{Introduction}

The study of quasi two-dimensional (2D) materials with frustrated magnetic
exchange interactions is a field of intense research. This research is
driven by the quest for systems which may exhibit exotic magnetic phases
instead of simple long range anti/ferromagnetic order (AFM/FM LRO)
\cite{FMBook}. Prominent examples of such phases are spin liquids (SL),
with no ordering of any type, or valence bond states. The latter may occur
as solids (VBS) with no breaking of lattice symmetries but potentially
other hidden order, such as eg. string ordering, moreover valence bond
crystals (VBC) are frequent, where lattice symmetries are directly broken
in favor of eg. columnar or plaquette ordering
\cite{FMBook,QMBook,FisherRev}.
\begin{figure}[t]
\includegraphics[width=5cm]{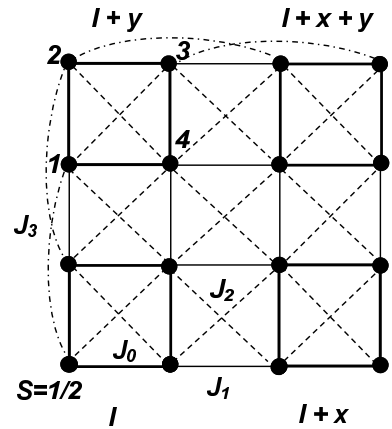}
\caption{The \emph{generalized} $J_1$-$J_2$-$J_3$ model considered in this
work. Solid circles represent spin$-1/2$ moments. Plaquettes (bold solid
lines) are non-locally coupled by nearest ($J_1$), next nearest ($J_2$) and
next-next nearest ($J_3$) interactions, represented by thin solid, dashed
and dot-dashed lines, respectively. For clarity only some of the $J_3$
couplings are depicted. On each isolated plaquette, the couplings along the
square edges are $J_0$ (bold solid lines) and across the diagonals $J_2$.
At $J_1=J_0$ the $J_1$-$J_2$-$J_3$ model is recovered. $J_0$ couplings are
set to unity hereafter.} \label{lattice}
\end{figure}
Even for simple frustrated systems a quantitative understanding of the
complete phase diagram is still lacking. The AFM spin-$1/2$ $J_1$-$J_2$
model on the square lattice, is a paradigmatic case in this respect. This
model corresponds to Fig.\ref{lattice}, for $J_1=J_0>0$, $J_2>0$ and
$J_3=0$, where $J_1$ (and $J_0$) is the nearest neighbor exchange
interaction and frustration is induced by the next-nearest neighbor
exchange interaction $J_2$. Experimentally, Li$_2$VOXO$_4$ (X $=$ Ge, Si),
which has been discovered recently, is a promising candidate to realize
the $J_1$-$J_2$ model in the range $J_2/J_1 \sim 5-10$ \cite{Melzi00,
Rosner02}. Two limiting cases of $J_1$-$J_2$ model are well-understood. For
$J_2 = 0$ and $J_1> 0$, the system is the 2D Heisenberg AFM, which exhibits
N\'eel LRO. In the opposite limit, $J_1/J_2 \rightarrow 0$, the system
turns into a set of two decoupled AFMs on the 2D A and B sublattices. These
lock into a collinear state by the order-from-disorder effect due to the
finite $J_2$. In the intermediate regime, $0.4 \approx (J_2/J_1)_{c_1}
< J_2 / J_1 < (J_2/J_1)_{c_2} \approx 0.6$, both, the N\'eel
and the collinear state are know to be unstable. Here, different
approaches, including exact diagonalization (ED) \cite{Dagotto99,
Poilblanc91, Schulz96}, quantum Monte Carlo (QMC) \cite{Capriotti00,
Capriotti01}, spin wave theory (SW) \cite{Chandra88}, large$-$N expansion
\cite{Read89}, and series expansion (SE) \cite{Gelfand90, Singh99, Kotov00,
Sushkov01, Sushkov02, Sirker06}, have confirmed that one or several
quantum disordered phases with a singlet ground state and a gap to
magnetic excitations may be present. The precise nature of the
intermediate phase (S), however, is still controversial. In
the simplest scenario, considering the existence of a single intermediate
phase only, a plaquette VBC \cite{Capriotti00}, a columnar VBC
\cite{Read89} and
a SL \cite{Capriotti01} have been proposed. Other studies suggest that the
intermediate phase could be composed of two SL-like phases
\cite{Sushkov01}.

The main purpose of this paper is to put the $J_1$-$J_2$ model into a
broader perspective, by considering an extended version, i.e. the
$J_1$-$J_2$-$J_3$ model, which is depicted in Fig.\ref{lattice} for $J_1 =
J_0$ and includes a third nearest-neighbor interaction $J_3$ (for clarity
only some of the $J_3$ couplings are shown). Classically, the competing
interactions $J_2/J_1$ and $J_3/J_1$ lead to four ordered phases of the
$J_1$-$J_2$-$J_3$ model \cite{Moreo90, Chubukov91, Ferrer93}. Among them,
N\'eel and helicoidal phases, which are separated by a classical critical
line $(J_2 + 2 J_3)/J_1 = 1/2$ exist.  The N\'eel phase remains rather
stable against quantum fluctuations, although it has been conjectured that
critical line, at $J_2=0$, should be shifted to $J_3 / J_1 > 1/4$ once the
quantum model is considered \cite{Ferrer93}.

The nature of the quantum phases in a selected region $J_1$, $J_2$, and
$J_3$ has been considered recently by Mambrini {\it et al.}
\cite{Mambrini06}. By employing ED and diagonalization in a subset of
short-range valence bond singlets (SRVB method) these authors have found
evidence for a VBC ordered, gapped plaquette phase in an extended region
around the line $(J_2 + J_3)/J_1 = 1/2$ and $J_3 \geq J_2$. In this paper
we will complement and extend these findings by performing SE analysis. In
particular we will aim at a quantitative determination of the extension of
the plaquette phase around the previously mentioned line by localizing the
critical lines for a closure of the triplet gap.

Our strategy will be to analyze perturbatively the evolution of the ground
state of a \emph{generalized} version of the $J_1$-$J_2$-$J_3$ model. For
this version $J_0\neq J_1$. At $J_{1,3}=0$ and $J_2\neq 0$ only on those
squares formed by the $J_0$-links the generalized $J_1$-$J_2$-$J_3$ model
shown in Fig.\ref{lattice} exhibits a product-state of disconnected bare
four-spin 'plaquettes'. This will be the unperturbed ground state from
which we start. The local $J_0$ couplings (bold lines) will be set to unity
hereafter. Therefore at $J_1=1$ we recover the $J_1$-$J_2$-$J_3$ model (in
units of $J_1$) from the generalized model. The Hamiltonian of generalized model
is
\begin{equation}\label{H}
  H = H_0 + V; \hspace{0.1cm}
  H_0 = \sum_{\mathbf{l}} h_{0, \mathbf{l}}; \hspace{0.1cm}
  V = \sum_\mathbf{l} \left( V_{1,\mathbf{l}} + V_{2,\mathbf{l}}
  + V_{3,\mathbf{l}} \right),
\end{equation}

\noindent where $h_{0, \mathbf{l}}$ refers to the local plaquette at site $\mathbf{l}$,
given by

\begin{eqnarray}
  h_{0, \mathbf{l}} &=& [\mathbf{S}_1 \cdot \mathbf{S}_2 + \mathbf{S}_2 \cdot \mathbf{S}_3
    + \mathbf{S}_3 \cdot \mathbf{S}_4 + \mathbf{S}_4 \cdot \mathbf{S}_1   \\
   & &  + J_2 ( \mathbf{S}_1 \cdot \mathbf{S}_3 + \mathbf{S}_2 \cdot \mathbf{S}_4 )]
   _\mathbf{l} \nonumber \\
  &=&  \frac{1}{2} [ \mathbf{S}_{1234}^2 - \mathbf{S}_{13}^2 -
  \mathbf{S}_{24}^2
     + J_2 ( \mathbf{S}_{13}^2 + \mathbf{S}_{24}^2 - 3 ) ]_\mathbf{l},\nonumber
\end{eqnarray}

\noindent in which $\mathbf{S}_{1 \ldots n}=\mathbf{S}_1 + \ldots + \mathbf{S}_n$.
$V_{1,\mathbf{l}}$, $V_{2,\mathbf{l}}$ and $V_{3,\mathbf{l}}$ in Eq.(\ref{H}) represent
the inter-plaquette coupling at site $\mathbf{l}$ via nearest ($J_1$), next nearest ($J_2$)
and next-next nearest ($J_3$) interactions, respectively.

\begin{eqnarray}
   V_{1,\mathbf{l}} &=& J_1 [ \mathbf{S}_{3,\mathbf{l}} \cdot \mathbf{S}_{2, \mathbf{l}+ \mathbf{x}}
   + \mathbf{S}_{4,\mathbf{l}} \cdot \mathbf{S}_{1, \mathbf{l}+ \mathbf{x}} \\
&& + \mathbf{S}_{2,\mathbf{l}} \cdot \mathbf{S}_{1, \mathbf{l}+ \mathbf{y}}
       + \mathbf{S}_{3,\mathbf{l}} \cdot \mathbf{S}_{4, \mathbf{l}+ \mathbf{y}}]; \nonumber \\
V_{2,\mathbf{l}} &=& J_2 [ \mathbf{S}_{4,\mathbf{l}} \cdot \mathbf{S}_{2, \mathbf{l}+ \mathbf{x}}
   + \mathbf{S}_{3,\mathbf{l}} \cdot \mathbf{S}_{1, \mathbf{l}+ \mathbf{x}} \nonumber \\
&& + \mathbf{S}_{2,\mathbf{l}} \cdot \mathbf{S}_{4, \mathbf{l}+ \mathbf{y}}
   + \mathbf{S}_{3,\mathbf{l}} \cdot \mathbf{S}_{1, \mathbf{l}+ \mathbf{y}} \nonumber \\
&& + \mathbf{S}_{3,\mathbf{l}} \cdot \mathbf{S}_{1, \mathbf{l}+ \mathbf{x}+ \mathbf{y}}
   + \mathbf{S}_{4,\mathbf{l}+ \mathbf{y}} \cdot \mathbf{S}_{2, \mathbf{l}+ \mathbf{x}}]; \nonumber \\
V_{3,\mathbf{l}} &=& J_3 \sum_{i=1}^{4} \left( \mathbf{S}_{i,\mathbf{l}}
\cdot \mathbf{S}_{i, \mathbf{l}+ \mathbf{x}}
+ \mathbf{S}_{i,\mathbf{l}} \cdot \mathbf{S}_{i, \mathbf{l}+ \mathbf{y}}\right). \nonumber \label{V}
\end{eqnarray}

\noindent Table \ref{tableH0} shows the eigenstates of a local plaquette Hamiltonian, $h_{0, \mathbf{l}}$,
in which each state is labeled by the ground state energy: $e_0$, the total spin: $S_{1234}$, and the spin
along each diagonal: $S_{13}$ and $S_{24}$.

\begin{table}[t]
\begin{tabular}{c|c|c|c|c}
\hline    state & $q\equiv e_0 + 2$ & $S_{1234}$ & $S_{13}$ & $S_{24}$
\\ \hline
$|{\cal S}_t\rangle$ & $\frac{1}{2}J_2$ & 0 & 1 & 1 \\
$|{\cal T}_t\rangle$ &  $\frac{1}{2}J_2 + 1$ & 1 & 1 & 1 \\
$|{\cal S}_s\rangle$ & $ -\frac{3}{2}J_2 + 2$  & 0 & 0 & 0  \\
$|{\cal T}_{ts}\rangle$ & $ -\frac{1}{2}J_2 + 2$ & 1 & 1 & 0 \\
$|{\cal T}_{st}\rangle$ & $ -\frac{1}{2}J_2 + 2$ & 1 & 0 & 1 \\
$|{\cal Q}_t\rangle$ & $ \frac{1}{2}J_2 + 3$ & 2 & 1 & 1 \\
\hline
\end{tabular}
\caption{Eigenstates of a local plaquette Hamiltonian, $h_{0, \mathbf{l}}$.
Each state is labeled by the energy: $e_0$ and the quantum numbers: $S_{1234}$,
$S_{13}$ and $S_{24}$. Note that $J_0$ coupling has been set to unity.} \label{tableH0}
\end{table}

\noindent From this table it follows that for $0 \leq J_2 < 1$ the ground
state is $|{\cal S}_t\rangle$, i.e. a spin singlet along the plaquette and
triplets along  the diagonals. For $0 \leq J_2 < \frac{1}{2}$ the first excited
state is $|{\cal T}_t\rangle$, i.e. triplets along both the plaquette and the diagonals.
At $J_2 = 1$ there is a crossover in the ground state energy and thereafter
the ground state is $|{\cal S}_s\rangle$, i.e. singlets along the plaquette and
the diagonals. The other states are total triplets, $|{\cal T}_{ts}\rangle$ and
$|{\cal T}_{st}\rangle$, consisting of a triplet on one of the diagonals and a triplet
on the other one. Finally, there is a quintet state, $|{\cal Q}_t\rangle$.

\section{Series expansion by continuous unitary transformation}

In this Section we briefly describe the SE expansion in terms of
$J_1$, $J_2$ and $J_3$. First, we rewrite the Hamiltonian (Eq.(\ref{H})) as
\begin{equation}\label{Hrew}
    H=H_0(J_2=0) + J_2 O_{2}^{0}+
    \sum_{i=1}^3 \left( J_i \sum_{n=-N}^N   O_i^n \right),
\end{equation}
where $H_0$ has been split into the first two terms.
The first one, $H_0(J_2=0)$, has a set of equally spaced
energy levels (Table \ref{tableH0}). These are labeled with
a \emph{total particle-number} operator:
$Q=\sum_{\mathbf{l}}q_{\mathbf{l}}(J_2=0)$. $Q=0$ corresponds
to zero particle states:
$|\mathbf{0} \rangle \equiv \prod_\mathbf{l}|{\cal S}_t\rangle_\mathbf{l}$.
$Q=1$ sector corresponds to one-particle states:
$|\mathbf{1}\rangle_\mathbf{l^\prime} \equiv |{\cal T}_t\rangle_\mathbf{l^\prime}
\bigotimes \prod_{\mathbf{l}\neq\mathbf{l^\prime}} |{\cal S}_t\rangle_\mathbf{l} $,
i.e., a local triplet at site $\mathbf{l^\prime}$ created from the vacuum.
$Q \geq 2$ sector of the spectrum is of multiparticle nature.

The second term in Eq.(\ref{Hrew}) refers to local contributions in $H_0$
proportional to $J_2$. The last three terms in the same Eq.(\ref{Hrew})
represent the inter-plaquette interactions, via $J_1$, $J_2$ and $J_3$,
respectively. There, $O_i^n$ operators non-locally create ($n \geq 0$)
and destroy ($n < 0$) quanta within the ladder spectrum of $H_0(J_2=0)$.
The explicit tabulation of $O_i^n$ in this model shows that $N \leq 4$  \cite{Oelem}.

It has been shown \cite{Knetter00} that models of type Eq.(\ref{Hrew})
allow for SE by means of Wegner's continuous unitary transformation (CUT) method
\cite{Wegner94}. The basic idea is to map $H \rightarrow H_{\mathrm{eff}}$,
where
\begin{equation}\label{Heff}
    H_{\mathrm{eff}} = H_0 + \sum _{k, m, l=1}^\infty  C_{k,m,l} J_1^k
J_2^m J_3^l.
\end{equation}
The $C_{k,m,l}$ operators in Eq.(\ref{Heff}) involve products of the
$O_i^n$ operators of Eq.(\ref{Hrew}). However, as the main point and unlike
in $H$, the effective Hamiltonian $H_{\mathrm{eff}}$ is constructed to have
a block diagonal structure, where each block has a {\em fixed} number of
particles $Q$ of $H_0(J_2=0)$. This is achieved order by order in the
expansion. We refer to Ref.\cite{Knetter00} for further details. In the
following Sections we will apply this technique to calculate the ground
state energy and the one-particle excitations.

\section{Dispersion of one-triplet excitations}
In this Section we evaluate the dispersion of one-triplet states for
different values of the coupling constants, $J_1$, $J_2$ and
$J_3$. To this end, it is necessary to diagonalize $H_{\mathrm{eff}}$ in
the $Q=1$ sector of $H_0(J_2=0)$, i.e., the subspace spanned by
$|\mathbf{1}\rangle_{\mathbf{l}}$ states. Q-conservation implies that the
sole action of $H_{\mathrm{eff}}$ on the local triplet states refers to
translation in real space, i.e.,
\begin{equation}\label{HeffTShift}
    H_{\mathrm{eff}}|\mathbf{1}\rangle_{\mathbf{0}}=
    \sum_{\mathbf{l}} c_{\mathbf{l}} |\mathbf{1}\rangle_{\mathbf{l}},
\end{equation}
\noindent where the $c_{\mathbf{l}}$'s are the hopping amplitudes of
a local triplet from origin $\mathbf{0}$ to site $\mathbf{l}$.
\noindent Due to the lattice translational invariance Eq.(\ref{HeffTShift})
can be diagonalized by Fourier transformation
\begin{equation}\label{E1trip}
    E_{\mathbf{1}}(\mathbf{k})=
    \sum_{\mathbf{l}} c_{\mathbf{l}} \exp(i \mathbf{k} \cdot \mathbf{l}).
\end{equation}
\noindent From this, the dispersion $\omega(\mathbf{k})$ follows as
\begin{equation}\label{dispF}
    \omega(\mathbf{k})= E_{\mathbf{1}}(\mathbf{k}) - E_{\mathbf{0}},
\end{equation}
\noindent where the ground state energy, $E_{\mathbf{0}}$,
is obtained by applying Q-conservation  to the 0-particle sector,
i.e.,  $E_{\mathbf{0}}=\langle 0| H_{\mathrm{eff}}|0\rangle$.
It is important to note that, even without an explicit discussion of this
quantity, Eq.(\ref{dispF}) requires a full calculation of the
ground state energy up to the same order as the hopping
amplitudes.

By symmetry considerations not all the $c_{\mathbf{l}}$'s
are independent, which leaves only a subset of them to be calculated.
Usually, in CUT applications, the $c_{\mathbf{l}}$'s at $O(n)$ are obtained
in the thermodynamic limit, by considering finite clusters which are large enough to
embed all the paths of length $n$ \cite{step-note} that connect origin
$\mathbf{0}$ with site $\mathbf{l}$. In our model, due to the number of
couplings considered and its dimensionality, this method becomes computationally
very demanding.
Alternatively, we have implemented a linked cluster approach, with subgraph
subtraction to obtain the $c_{\mathbf{l}}$'s. We refer to Ref.\cite{SEBook}
for technical details of this method.

We have evaluated {\em analytic} expressions for the triplet
dispersion, $\omega(\mathbf{k})$, keeping all
three independent variables $J_1$, $J_2$, and $J_3$, i.e. without
any parametrization, up to $O(5)$ \cite{disp-expl}.

\begin{figure}[t]
\vspace{0.2cm}\includegraphics[angle=-90,width=8.5cm]{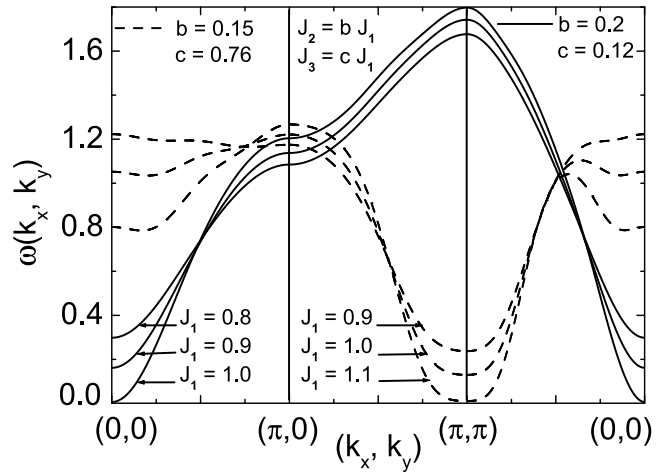}
\caption{Triplet dispersion $\omega(\mathbf{k})$ as a function of the
wave vector $\mathbf{k}=(k_x,k_y)$ along the path
$\mathbf{k}=(0,0)-(\pi,0)-(\pi,\pi)-(0,0)$. Two families of curves in
coupling-constant space with parametrization $J_2=b J_1$ and $J_3=c J_1$
close to the actual $J_1$-$J_2$-$J_3$ model at $J_1=1$ have been selected
to show triplet softening, i.e. $\omega(\mathbf{k})=0$. The instability at
$\mathbf{k_{c1}}=(0,0)$ occurs at small values of $J_3$, with respect to
$J_1$ (solid lines). For larger values of $J_3$ the instability is at
$\mathbf{k_{c2}}=(\pi,\pi)$ (dashed lines).} \label{disp}
\end{figure}

Fig.\ref{disp} shows the dispersion obtained at $O(5)$, as a function of
wave vector \textbf{k}, along high symmetry directions and for different
values of the couplings. We have chosen paths in the couplings space that
show the instabilities of the plaquette phase associated with triplet
softening, i.e. $\omega(\mathbf{k})=0$. We have selected two families of
curves, parametrized according to $J_1$, $J_2=b J_1$, and $J_3=c J_1$,
around $J_1=1$, the latter being the point where the $J_1$-$J_2$-$J_3$
model (in units of $J_1$) is recovered (see Fig.\ref{lattice}).

As shown in Fig.\ref{disp}, triplet softening occurs at a critical wave
vector of $\mathrm{\mathbf{k}_{c1}}=(0,0)$ for the specific value of $(J_1,
J_2, J_3) \approx (1, 0.2, 0.12)$, i.e. for relatively small values of
$J_3$, as compared to $J_2$ (solid lines). Additionally, for larger values
of $J_3$ a critical wave vector at $\mathrm{\mathbf{k}_{c2}}=(\pi,\pi)$
(dashed lines) can be observed for the particular value of $(J_1, J_2, J_3)
\approx (1.1, 0.15, 0.76)$ (dashed line). We have found no other values for
critical wave vectors. Fig.\ref{disp} clarifies the type of critical
points that have to be expected and is a first indication of the
qualitative relevance of $J_3$ on the possible ground states of the model.
To obtain a quantitative picture, the stability region of the plaquette
phase in $J_1$-$J_2$-$J_3$ space will be studied in detail in the following
Sections.

\section{Stability of the plaquette phase}

\begin{figure}[t]
\hspace{0.5cm}
\includegraphics[angle=-90,width=8.5cm]{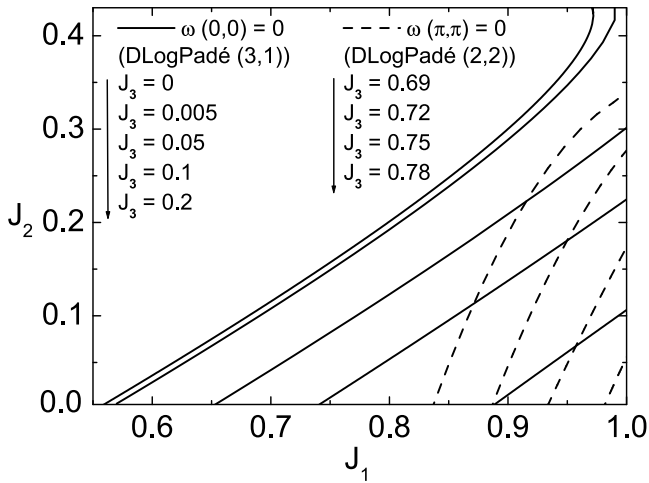}
\caption{Critical lines $(\omega(\mathbf{k}_c)=0)$ in $J_1$-$J_2$ plane and
$J_3$ as parameter. Triplet softening occurs at $\mathbf{k}_{c1}=(0,0)$ and
$\mathbf{k}_{c2}=(\pi,\pi)$, shown with solid and dashed lines,
respectively. In all cases results from Dlog-Pad\'e analysis are depicted.
For $0 \leq J_3 \lesssim 0.4 $, the instability at $\mathbf{k}_{c1}$ limits
the plaquette phase, projected onto $J_1-J_2$ plane. But only for $J_3
\gtrsim 0.05$ a plaquette phase appears in $J_1$-$J_2$-$J_3$ model
(critical lines cross $J_1=1$). In particular, for $J_1=1$, $J_2$, and
$J_3=0$, i.e. in the $J_1$-$J_2$ model, the plaquette phase is not present.
For $0.7 \lesssim J_3 \lesssim 0.8$, the critical lines at
$\mathbf{k}_{c2}$ limit the plaquette phase projected onto $J_1-J_2$
plane.}
\label{gap00PiPiJ1J2J3p}
\end{figure}

In this Section we discuss the quantum critical lines, resulting from the
closure of the plaquette triplet gap, which resembles second order quantum
phase transitions. This will give us a quantitative estimate of the
stability region of the plaquette phase. In particular we are interested in
a possible adiabatic connection of the isolated bare plaquette phase (with
only local $J_2 \neq 0$) up to the value of $J_1=1$. This analysis is shown
in Fig.\ref{gap00PiPiJ1J2J3p} which depicts the borders of the stability
region projected onto $J_1-J_2$ plane, taking $J_3$ as parameter.

Fig.\ref{gap00PiPiJ1J2J3p} displays two families of critical lines,
corresponding to the closure of the triplet gap
$\omega(\mathrm{\mathbf{k}_c})=0$ for $\mathrm{\mathbf{k}_{c1}}=(0,0)$ and
$\mathrm{\mathbf{k}_{c2}}=(\pi, \pi)$, with dotted and solid lines,
respectively. First, it is obvious that independently of $J_2$ and $J_3$
the plaquette phase extends from the origin, $J_1=J_2=J_3=0$ (not shown in
Fig.\ref{gap00PiPiJ1J2J3p}) up to $J_1 \approx 0.55$ below which there are
no signals of triplet softening. Second, we focus on the
$\omega(\mathrm{\mathbf{k}_{c1}})=0$ instability. In the case of $J_3=0$,
as can be observed from the Figure, the critical line almost reaches, but
does not cross the line $J_1=1$. In other words: the $J_1$-$J_2$ model does
not show a plaquette phase. This result is consistent with the previous SE
analysis on $J_1$-$J_2$ model in Ref.\cite{Singh99}.

\begin{figure}[t]
\vspace{0.15cm}\hspace{-0.3cm}
\includegraphics[width=6.7cm]{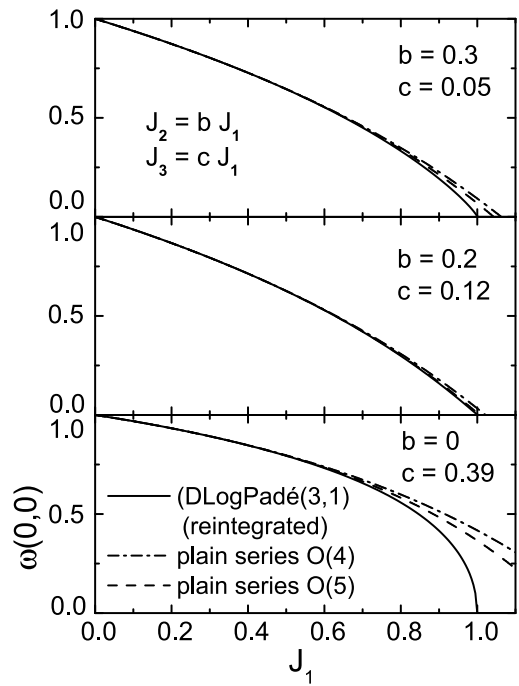}
\caption{Comparison between the triplet gap at $\mathbf{k}_{c}=(0,0)$
obtained by means of plain series and a particular Dlog-Pad\'e approximant
along selected straight line paths in couplings space, for a case in which
the triplet gap closes at $J_1=1$. Results from reintegrated
Dlog-Pad\'e(3,1), plain series at $O(4)$ and $O(5)$ are shown with solid,
dot-dashed and dashed lines, respectively. For $J_1 \lesssim 0.5$ the
agreement is very good in all cases. Closer to $J_1$-$J_2$-$J_3$ model,
i.e. to $J_1=1$, clear differences between the Dlog-Pad\'e and the plain
series arise.} \label{DLP}
\end{figure}

Third, we consider the simultaneous effect of $J_1$, $J_2$ and $J_3$. As
can be observed in Fig.\ref{gap00PiPiJ1J2J3p}, increasing the values of
$J_3$ enlarges the region of stability of
the plaquette phase in the $J_1-J_2$ plane in terms of the critical line
$\mathrm{\mathbf{k}_{c1}}=(0,0)$ (solid lines). Most important, finite
$J_3$ helps to stabilize the plaquette phase at $J_1=1$. In fact,
already for $J_3 \approx 0.05$ the critical line crosses $J_1=1$.
For $J_3 \approx 0.4$ the solid critical line merges with the lower
righthand corner of Fig.\ref{gap00PiPiJ1J2J3p} and the plaquette phase
extends over all of the $J_1-J_2$ plane shown. These results are consistent
with Ref.\cite{Mambrini06}.

Now we turn to the plaquette phase stability region, projected onto
$J_1-J_2$ plane, limited by the critical lines
$\omega(\mathrm{\mathbf{k}_{c2}})=0$ (dotted lines in
Fig.\ref{gap00PiPiJ1J2J3p}). We find a similar tendency as for
$\mathbf{k}_{c2}$, i.e. the region of stability of the plaquette phase in
the $J_1-J_2$ plane is enlarged by increasing $J_3$. In this case however
the impact of $J_3$ is somewhat less significant as compared to $J_2$.

Technically, the critical lines of Fig.\ref{gap00PiPiJ1J2J3p} have been
obtained using Dlog-Pad\'e analysis, rather than the plain series. This is
known to improve the accuracy of locating the critical points
significantly. For details on this technique we refer the reader to the
literature \cite{Guttmann89}. In order to work with single variable
Dlog-Pad\'es we have scanned the exchange coupling space by means of
straight lines, parametrized according to $(J_1, J_2=b J_1, J_3=c J_1)$.
For fixed values of $b$ and $c$ this amounts to a single variable, i.e.
$J_1$.

To assess the impact of the Dlog-Pad\'e analysis, we show its result for
$\omega(\mathrm{\mathbf{k}_{c1}})$, for a particular Dlog-Pad\'e
approximant and a case in which the triplet gap closes at $J_1=1$
(Fig.\ref{DLP}). A Similar analysis has been done for all the critical
lines calculated, including several Dlog-Pad\'e approximants in each case.
In this Figure, the solid line refers to the reintegrated Dlog-Pad\'e
$(3,1)$, and the dot-dashed and dashed lines show the plain series at
O$(4)$ and O$(5)$, respectively. From there it is clear that for $J_1
\lesssim 0.5$ the agreement between the reintegrated Dlog-Pad\'e and the SE
at O$(4)$ and O$(5)$ is very good. In fact, all plots are indistinguishable
on the scale used. This provides a qualitative measure of the convergence
of the series. For $J_1 \gtrsim 0.5$ and closer to criticality (at $J_1=1$
in this case) however, we rely on the Dlog-Pad\'e technique in order to
describe the closure of the gap.

\section{Plaquette phase at $J_1=1$}
Here we analyze the extent of the plaquette phase on the $J_2$-$J_3$ plane
at $J_1=1$, i.e. for the actual $J_1$-$J_2$-$J_3$ model, written in units
of $J_1$. As it was mentioned in the Introduction, ED calculations for
$J_1$-$J_2$-$J_3$ model, using the complete Hilbert space and a restricted
space of short-range dimer singlets, provide strong evidence for the
existence of a plaquette phase around the line $J_2+J_3=1/2$ at $J_1=1$, in
particular for $J_3 \geq J_2$ \cite{Mambrini06}. In this Section, we will
extend this study by specifying the extension of this phase as obtained
from SE. To this end, we proceed as in the previous Section, i.e. the
critical lines are obtained by analyzing the closure of the triplet gap,
i.e. solutions of $\omega(\mathbf{k})=0$.

\begin{figure}[t]
\vspace{0.2cm}
\includegraphics[angle=-90,width=8.5cm]{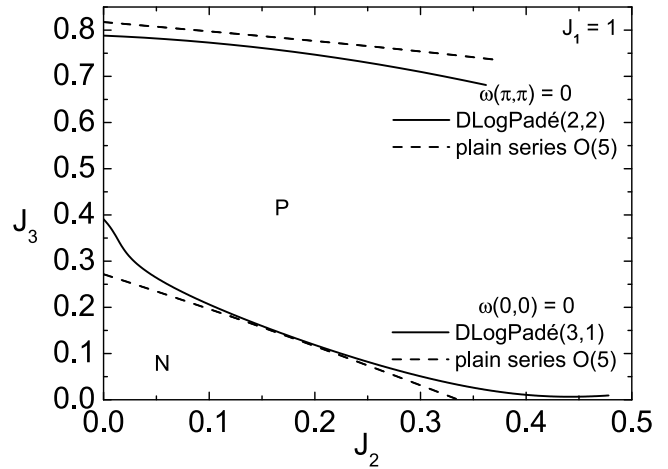}
\caption{Extension of the plaquette phase in $J_2-J_3$ plane at $J_1=1$
(actual $J_1$-$J_2$-$J_3$ model). The lower and
upper critical lines represent the closure of the triplet gap at
$\mathbf{k}_{c1}=(0,0)$ and $\mathbf{k}_{c2}=(\pi,\pi)$, respectively,
which limit the plaquette phase (intermediate region labeled with P).
Solid and dashed lines represent the results obtained employing Dlog-Pad\'e
approximants and plain series, respectively. The plaquette phase extends
considerably around the straight line which connects $(J_2=0, J_3=0.5)$
with $(J_2\approx J_3, J_3\approx0.25)$, previously studied in
Ref.\cite{Mambrini06}. Unlike the plain series, our Dlog-Pad\'e analysis
suggests that the plaquette phase is not present in $J_1$-$J_2$ model
$(J_3=0)$ for the parameters studied.
However, for $0.35 \lesssim J_2 \lesssim 0.6$ the critical line is too
close to $J_3=0$ to allow for definite statements at this order of SE.}
\label{PPJ2J3J11}
\end{figure}

In Fig.\ref{PPJ2J3J11} we show the corresponding results.
The lower and upper critical lines mark the triplet softening at
$\mathbf{k}_{c1}$ and $\mathbf{k}_{c2}$, respectively, and enclose
the region of a finite triplet gap. I.e. this region refers to the
plaquette phase, labeled by 'P'. In this Figure, Dlog-Pad\'e approximants
are depicted by solid lines, and the results obtained by employing the
$O(5)$ plain series by dashed lines. We note that the critical lines
shown from Dlog-Pad\'e approximants in Fig.\ref{PPJ2J3J11} are
consistent with the pairs of ($J_{2c}, J_{3c}$) at $J_1=1$ shown in
Fig.\ref{gap00PiPiJ1J2J3p}.

Although, as in the previous Section, we base our results on the
Dlog-Pad\'e analysis, the agreement between the plain series and the
Dlog-Pad\'e approximants can be used to assess the convergence of the
series. From Fig.\ref{PPJ2J3J11}, it is clear that the best agreement for
the lower critical line is found in the intermediate region, i.e where
$0.1\lesssim J_{2,3} \lesssim 0.2$.

For the special case of $J_2=0$, i.e. for the pure $J_1$-$J_3$ model, it has
been conjectured that the classical critical line to N\'eel phase at
$J_3/J_1=0.25 $ ($J_1=1$ in our case) should be shifted to larger values in
the quantum model \cite{Ferrer93}. For all Dlog-Pad\'es analyzed, our
results confirm this conjecture, as e.g. for the $(3,1)$ Dlog-Pad\'e
approximant of the lower critical line shown in Fig.\ref{PPJ2J3J11}.

In conclusion we find that the plaquette phase extends considerably around
the straight line of maximal frustration, connecting $J_2=0, J_3=0.5$ with
$J_2=J_3\simeq 0.25$, which was studied in Ref.\cite{Mambrini06}. In
particular, as it can be seen in Fig.\ref{PPJ2J3J11}, the upper critical
line is rather far from the line of maximal frustration. Additionally, in
the limiting case $J_3=0$, we remain with the $J_1$-$J_2$ model. For the
latter, and as shown in the right lower corner of Fig.\ref{PPJ2J3J11}, and
unlike the plain series, the Dlog-Pad\'e analysis suggests that the
critical line does not intersect the $J_2$ axis. I.e. we find
no stability of the plaquette phase. This is in agreement with the SE
results of Ref.\cite{Singh99}. Yet, the proximity between the critical
line and the $J_2$ axis calls for caution on this finding with respect to
the convergence of the SE in this parameter range.

\section{Conclusions}
To summarize, using series expansion, based on flow equations we have
analyzed the zero temperature properties of the 2D spin-$1/2$
$J_1$-$J_2$-$J_3$ AFM. Starting from the limit of
decoupled plaquettes of a generalized $J_1$-$J_2$-$J_3$ model we have
evaluated three-parameter series up to $O(5)$ in the interplaquette
exchange couplings $J_1$, $J_2$ and $J_3$ for the ground state energy
and for the triplet dispersion.

We find a rather large range of $J_{1,2,3}$ couplings which adiabatically
connects to the state of isolated plaquettes and hosts a plaquette phase which is
stable against second order quantum phase transitions into magnetic states.
Our findings corroborate and enhance related predictions of Mambrini
\emph{et al.} \cite{Mambrini06} on the location of a stable
plaquette phase at $J_1=1 $.

For the particular case of the $J_1$-$J_2$ model at $J_3=0$, and consistently
with results obtained in Ref.\cite{Singh99}, our calculation predicts that
the plaquette phase is not stable in the parameter range which we have
investigated. However, higher order series expansions seem very desirable to
render such results more reliable. In particular, from our series we are
reluctant to draw any definite conclusions about the controversial region
$J_3=0$ and $0.4 \lesssim J_2 \lesssim 0.6$.

Finally, we emphasize that our analysis has been focused on the stability
of the plaquette phase with respect to second order transitions driven by
one-particle (triplet) excitations. Further instabilities, like first order
transitions or level crossings of excited states, other than elementary
triplets, could give rise to further reduction of the plaquette regime and
have not been considered here. Along this line, the two-particle sector,
which includes singlet excitations, may play a role that can be analyzed
using our SE technique. This deserves future investigation.

\section{Acknowledgments}
We would like to thank D. Poilblanc and A. L\"auchli for helpful
comments. This research was supported in part through
DFG Grant No. BR 1084/4-1.


\begin{thebibliography}{99}
\bibitem{Oelem} Explicit tabulation of $O_i^n$
is electronically available on request from the authors.
\bibitem{step-note} A graph of lenght $n$ is composed by $n$ steps.
Each step connects a given site with its nearest neighbors along
both $\mathbf{x},\mathbf{y}$ directions and the diagonals.
\end{thebibliography}
\end{document}